\documentclass[a4paper,11pt]{article}
\usepackage{pos}

\title{Upgrade and commissioning of the ALICE muon spectrometer}

\author{Livia Terlizzi, on behalf of the ALICE collaboration}

\affiliation{University and INFN Torino,\\
  Via Pietro Giuria 1, Torino, Italy}

\emailAdd{livia.terlizzi@cern.ch}

\abstract{...}

\FullConference{%
  FAIR next generation scientists - 7th Edition Workshop (FAIRness2022)\\
  23-27 May 2022\\
  Paralia (Pieria, Greece)
}

\begin{document}
\maketitle

\section{Introduction}

ALICE \cite{ALICEcollab} (A Large Ion Collider Experiment) at the CERN Large Hadron Collider (LHC) is
designed to study proton--proton and heavy--ion collisions at ultra-relativistic energies.
The main goal of the experiment is to assess the properties of quark--gluon plasma, a state
of matter where quarks and gluons are deconfined, which is reached in extreme conditions of
temperature and energy density \cite{QCD} \cite{QGP}. The production of quarkonia (${\rm c}\overline{\rm c}$ and ${\rm b}\overline{\rm b}$ bound states) is among the main observables to study the QGP.
During the ongoing Long Shutdown 2 (LS2) of the LHC (2019-2021), ALICE underwent a major upgrade of its apparatus, in view of the LHC Run 3, which started in 2022. The upgrade will allow a new ambitious program of high-precision measurements.
However, the detectors will have to cope with an increased collision rate, which will go
up to 50 kHz in Pb--Pb collisions.
For the Muon Spectrometer (MS) ALICE is implementing new hardware and software solutions.
The installation of a new vertex tracker, the Muon Forward Tracker (MFT) upstream from the absorber in the acceptance of the MS will improve the current measurements and enable new ones. It will allow one to separate, for the first time in ALICE in the forward-rapidity region, the prompt (i.e.\ the ones directly produced in the interaction point) and non-prompt (i.e.\ the ones coming from beauty-hadron decays) contributions to the charmonium yield. The matching of the muon tracks reconstructed in the MFT with those in the Muon
Spectrometer will provide a precise determination of the track parameters in the vicinity of the interaction point allowing one to resolve the decay vertices of non-prompt charmonia in a broad interval of transverse momenta down to $p_{\rm T}$ = 0. It will also improve significantly the invariant mass resolution, allowing for a better separation of the J/$\psi$ and $\psi$(2S) states.
In addition, the front-end and readout electronics of the Muon Tracking system (Cathode
Pad - Cathode Strip Chambers) and of the Muon Identification system (Resistive Plate Chambers) has been
upgraded, in order to optimize the detector performance in the new running conditions.
A detailed description of the MS upgrades, together with the results from
the commissioning with cosmic rays and the first LHC beams, will be presented in this
talk.

\section{The ALICE Muon Spectrometer (MS)}

\begin{figure}[tb]
	\centerline{\includegraphics[width=9cm]{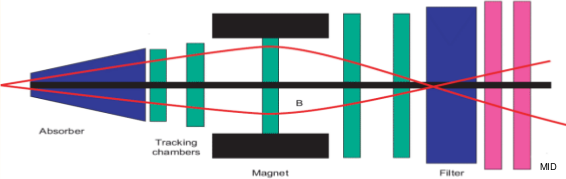}}
	\caption{Schematic side view of the Muon Spectrometer during Run 1 and Run 2.}
	\label{MSlayout}
\end{figure}

The ALICE Muon Spectrometer detects muons in the pseudorapidity range -4 $< \eta <$ -2.5 \cite{MSALICE}, which enables the reconstruction of quarkonia, open heavy-flavor hadrons, and W$^\pm$ and Z$^0$ bosons via their muonic decay channels.
During the LS2 of LHC, ALICE achieved a major upgrade of its apparatus to cope with the increased Pb--Pb collision rate foreseen for Run 3 (from 10 kHz in Run 2 to 50 kHz for Run 3), and to allow a new ambitious program of high-precision measurements. 
Before the upgrade the MS consisted of a front hadron absorber to stop most hadrons emitted in the MS acceptance, followed by a tracking system (Muon CHambers, MCH) made of 5 stations of 2 planes of Cathode Pad Chambers (CPC) and Cathode Strip Chambers (CSC), a dipole magnet providing an horizontal field perpendicular to the beam axis and parallel to the LHC radius, an iron wall to filter the residual background of hadrons, and finally the Muon TRigger (MTR), now Muon IDentifier (MID), made of 2 stations (each consisting of 2 planes) of Resistive Plate Chambers (RPCs) (figure \ref{MSlayout}). \\

\subsection{ALICE Muon Physics topics during Run 2}

In order to get insight into the QGP properties in the forward rapidity region, the muon physics program in ALICE included the study of: \\
- quarkonium production, which is one of the most powerful probes of the QGP. Quarkonium suppression and regeneration in the QGP are investigated; \\
- open heavy-flavor hadrons, which are created at the beginning of collisions (production by hard processes) and are sensitive probes of the medium created in the collision since they experience every phase of the medium evolution; \\
- single muons and dimuons from W$^\pm$ and Z$^0$ boson decays; \\
- low-mass dimuons. \\

\subsection{Muon Spectrometer limitations and upgrade motivations}

By analyzing the data samples collected with the MS during the LHC Run 1 and Run 2, the ALICE Collaboration could obtain and publish a large number of results, which provided new and important insights, especially into the production of quarkonia and their role as probes of the QGP. Nevertheless, there were still some limitations. \\
The multiple scattering in the front absorber smears the track information in the MCH, deteriorating the resolution on the muon production vertex. For this reason there is no separation of prompt J/$\psi$ and non-prompt J/$\psi$, which is an important source of information for beauty studies. Moreover there is no disentanglement of muons originating from open charm and open beauty hadron decays, without making assumptions relying on theoretical calculations. 
Also, there is the need to reduce the background from non-prompt sources, since at present there are significant statistical uncertainties, especially at low masses and low p$_T$, in single and di-muon analysis due to combinatorial background from muons originating from $\pi$ and K decays. 
Additionally, it is important to improve the dimuon invariant mass resolution for light neutral resonances, limited now by the poor resolution on the dimuon opening angle. 
Finally, a faster read-out is needed for the present MS detectors, in order to cope with the higher expected interaction rates in Pb--Pb collisions for Run 3 \cite{ALICEloi}.

\section{Muon Spectrometer Upgrades for Run 3}

In order to improve the resolution of the muon tracks in the vicinity of the interaction point, and consequentially to overcome the already mentioned limitations of the MS, a new silicon-based high spatial resolution Muon Forward Tracker (MFT) has been installed upstream from the front absorber \cite{ALICEMFTaddendum} \cite{MFTphysics}. 
For Run 3 ALICE aims at collecting 13 nb$^{-1}$ of Pb--Pb data, and to readout all Pb--Pb interactions up to the maximum LHC collision rate of 50 kHz (instantaneous luminosity L = 6$\times$10$^{27}$ cm$^{-1}$ s$^{-1}$). In order to do so, the readout electronics of the MID and MCH have been upgraded \cite{ALICEreadoutupgrade}.

\subsection{Muon IDentification (MID) upgrade}

\begin{figure}[ht] 
	\begin{minipage}[h]{0.53\linewidth}
		\includegraphics[width=.8\linewidth,height=.1\textheight]{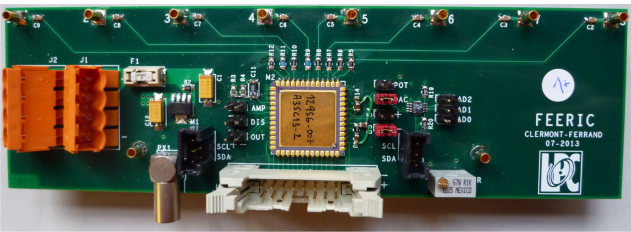} 
	\end{minipage}
	\begin{minipage}[h]{0.58\linewidth}
		\includegraphics[width=0.8\linewidth,height=.15\textheight]{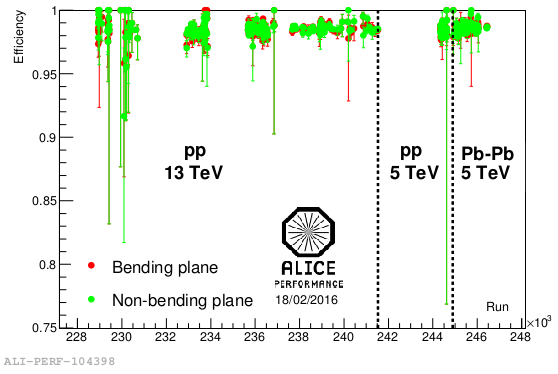} 
	\end{minipage} 
	\caption{Left panel: FEERIC front-end electronics board for the RPC MID. Right panel: efficiency of the MTR RPC equipped with FEERIC as a function of run number, measured in ALICE during Run 2.}
	\label{feeric_eff}
\end{figure}

Starting from Run 3, the ALICE experiment is running in continuous readout mode (i.e. without trigger) and the Muon Trigger has become a Muon IDentifier (MID). \\
MID consists of 72 Resistive Plate Chambers arranged in 2 stations of 2 planes each, equipped with orthogonal read-out strips (21k readout channels). 
The continuous readout mode required an upgrade of the readout electronics, which was completed during the Long Shutdown 2 \cite{ALICEreadoutupgrade}. Moreover, to cope with the increased counting rate and to reduce the RPC aging effects, the detectors will be operated at lower gain thanks to a new front-end electronics, called FEERIC ASIC (displayed in the picture in the left panel of Fig. \ref{feeric_eff}), which includes a pre-amplification stage of the signal. One RPC in the ALICE cavern was already equipped with FEERIC cards during Run 2. This RPC showed similar efficiency and performances with respect to the other RPCs equipped with the old front-end cards. The new FEERIC allows one to set a working point at a lower HV value with respect to Run 2. The efficiency is higher than 97$\%$ in both the bending and non-bending plane, for different collision systems (as it can be seen in the right panel of Fig.  \ref{feeric_eff}, where the efficiency is reported as a function of the run number for data taking periods with pp and Pb--Pb collisions). All the FEERIC cards have already been produced and installed during the Long Shutdown 2 and they are now under commissioning with first pp collision data.

\subsection{Muon CHambers (MCH) upgrade}

\begin{figure}[ht] 
	\begin{minipage}[h]{0.53\linewidth}
		\includegraphics[width=.9\linewidth,height=.12\textheight]{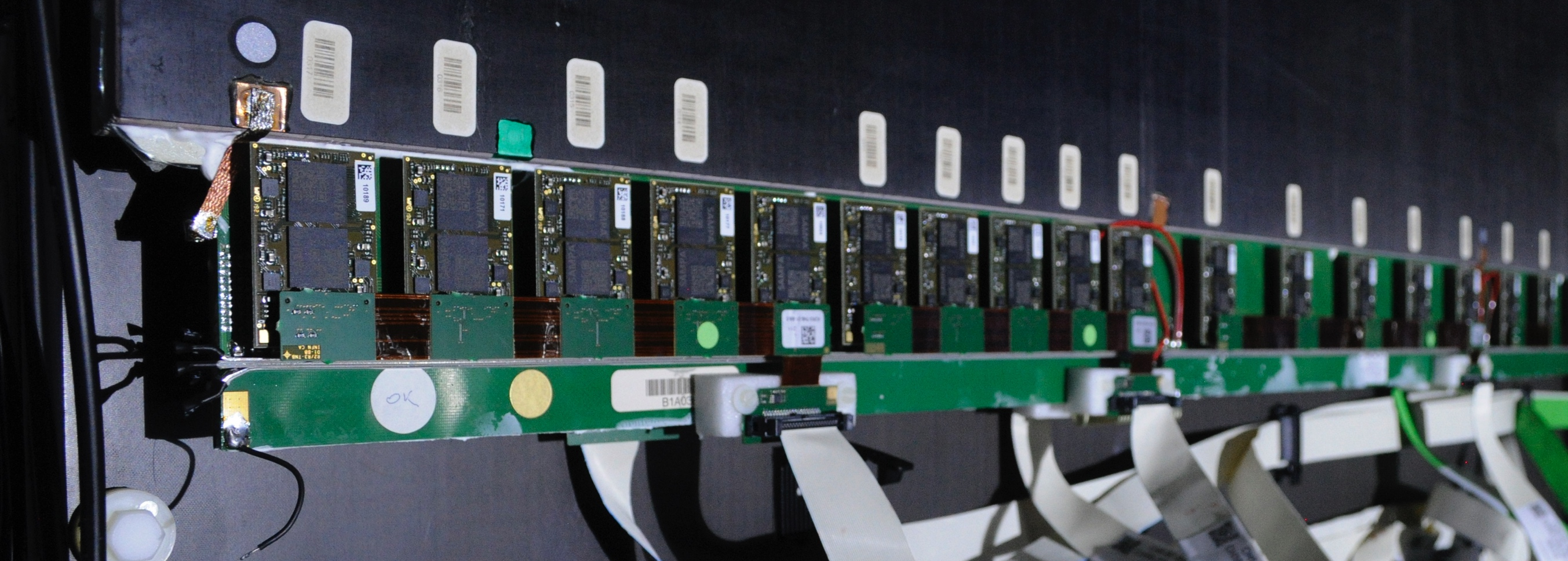} 
	\end{minipage}
	\begin{minipage}[h]{0.58\linewidth}
		\includegraphics[width=0.85\linewidth,height=.15\textheight]{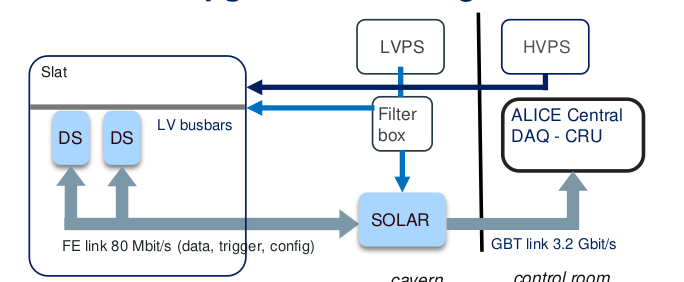} 
	\end{minipage} 
	\caption{Left panel: DUAL SAMPA front-end electronics board for the MCH. Right panel: new MCH readout chain for Run 3.}
	\label{sampa_readoutMCH}
\end{figure}

The muon tracking chambers are based on Cathode Pad Chambers (CPC) and Cathode Strip Chambers (CSC) arranged in 10 planes with 1.1 million readout channels and a spatial resolution of $\sim$ 100 $\mu$m. 
To cope with the higher collision rate in Run 3, the CPC and CSC detectors will have a new front-end electronics (FEE) and a new readout chain. The new FEE chip, called SAMPA, was developed jointly with the ALICE Time Projection Chamber upgrade project \cite{ALICEreadoutupgrade}. These new boards amplify, digitize, filter, and compress the signal. A picture of the board is shown in the left panel of Fig. \ref{sampa_readoutMCH}. 
The data from Dual Sampa (DS) boards (each hosting two SAMPA) are sent through FE links implemented on printed circuit boards to a new concentrator board called SOLAR, via GBT link protocol \cite{GBTlinks}. The SOLAR board collects data from several DS through FE links and sends them to the new Common Readout Unit (CRU). \\
The full chain is schematized in the right panel of Fig. \ref{sampa_readoutMCH}. It was successfully validated in a beam test in 2017 at the CERN SPS and all the components were installed during the LS2.

\subsection{Muon Forward Tracker (MFT)}

\begin{figure}[ht] 
	\begin{minipage}[h]{0.6\linewidth}
		\includegraphics[width=.9\linewidth,height=.18\textheight]{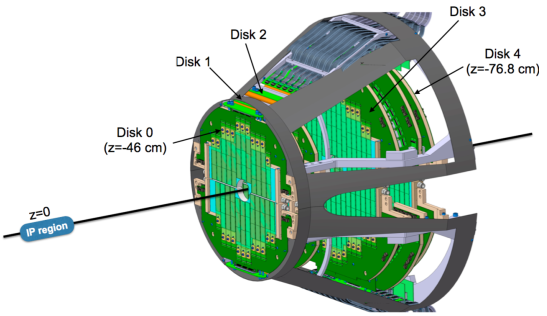} 
	\end{minipage}
	\begin{minipage}[h]{0.5\linewidth}
		\includegraphics[width=0.88\linewidth,height=.18\textheight]{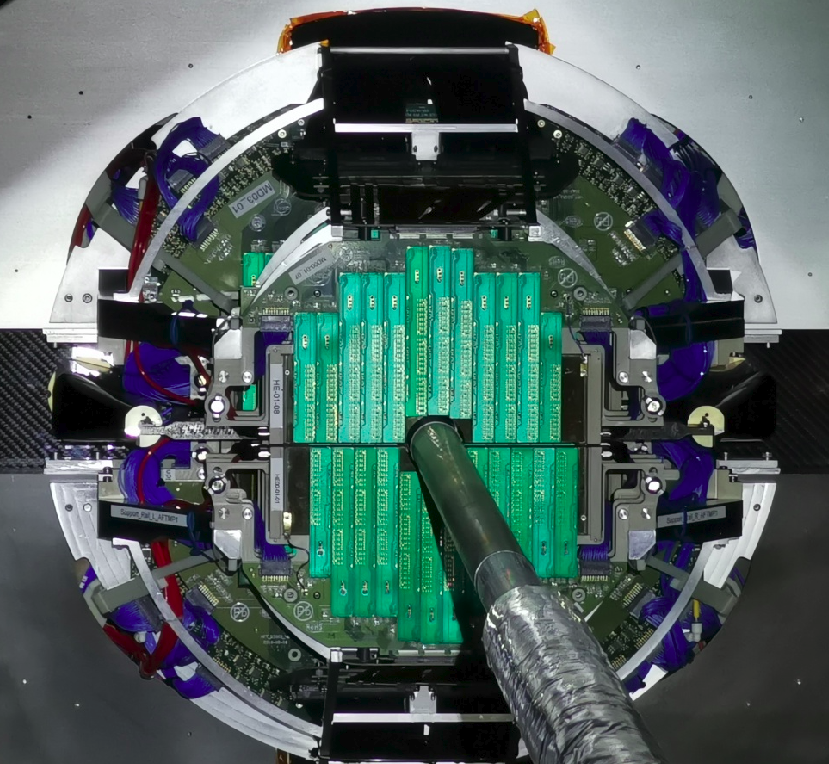} 
	\end{minipage} 
	\caption{Left panel: schematic layout of the MFT. Right panel: MFT installation in the ALICE cavern.}
	\label{MFT}
\end{figure}

The MFT is placed at forward rapidity and it will improve vertexing and tracking capabilities in the pseudorapidity range 2.5 $< \eta <$ 3.6, matching extrapolated muon tracks reconstructed in the MCH downstream of the absorber with MFT tracks upstream from the absorber. \\
It is constituted by 936 ALPIDE Silicon pixel sensors (0.4 m$^2$), based on Monolithic Active Pixel Sensors (MAPS), developed together with the new ALICE Inner Tracking System (ITS) \cite{ITSALICE}. The sensor size is 15 $\times$ 30 mm$^2$, while the pixel one is 27 $\mu$m $\times$ 29 $\mu$m. The MFT has a spatial resolution better than 5 $\mu$m and a cone structure formed by ten half disks positioned along the beam axis (as sketched in the left panel of Fig. \ref{MFT}). Each half-disk consists of structures called ladders, which host the ALPIDE chips (280 ladders of 2 to 5 sensors). It was installed in ALICE since January 2021. A picture taken during the installation phase is shown in the right panel of Fig. \ref{MFT}. It is now under commissioning with first pp collision data.

\subsection{New Physics measurements for Run 3}

\begin{figure}[tb]
	\centerline{\includegraphics[width=9cm]{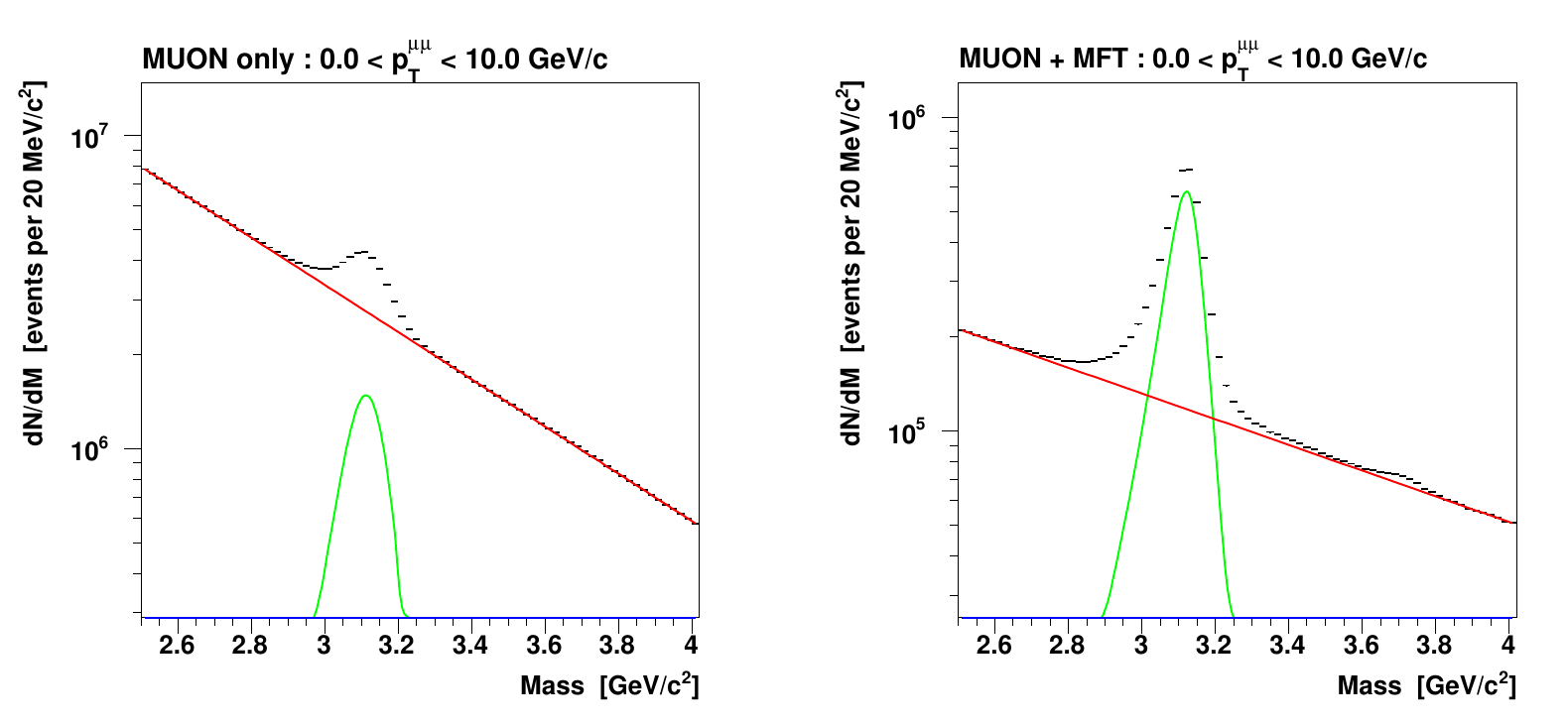}}
	\caption{J/$\psi$ and $\psi$(2S) peak at forward rapidity without (left) and with (right) the addition of the MFT, taken from~\cite{CERN-LHCC-2013-014}.}
	\label{charmonia_MFT}
\end{figure}

The MFT will allow signiﬁcant improvements and new measurements at forward rapidity concerning charmonium and open heavy flavor production, and low mass dimuons \cite{ALICEMFTaddendum}. \\
As a consequence of the improved resolution on the dimuon invariant mass, the MFT will improve the signal-to-background ratio by a factor of 5--6 with respect to the previous value in Run 2. This will allow for precise measurements of the J/$\psi$ and will reduce the uncertainties on $\psi$(2S) at forward rapidity in pp, p--Pb and Pb--Pb collisions. Dissociation and recombination models for charmonia can be tested by comparing the nuclear modification factors ($R_{\rm AA}$, defined as the ratio of the yield in Pb--Pb collisions and that in pp interactions scaled with the number of nucleon--nucleon collisions) of J/$\psi$ and $\psi$(2S) down to zero $p_{\rm T}$. Moreover precise measurement for the J/$\psi$ elliptic flow ($v_{\rm 2}$), which is the second coefficient of the Fourier expansion of the particle-momentum azimuthal-angle distribution, will also be possible \cite{MFTphysics}. 
The separation between prompt and non-prompt J/$\psi$ will be possible down to $p_{\rm T} = 0$ by measuring the pseudo-proper decay time associated to the secondary vertex. The separation is performed statistically by a simultaneous fit: the one on the invariant mass spectrum ﬁxes the normalization of the background and the inclusive J/$\psi$, while the one on the pseudo-proper decay time separates the non-prompt from the prompt J/$\psi$ component. This separation will give us access to beauty production and $R_{\rm AA}$ measurements down to zero $p_{\rm T}$ in central Pb--Pb collisions. Moreover it will serve as a useful tool to discriminate between diﬀerent models of charmonium regeneration in the QGP \cite{ALICEMFTaddendum}. 
Regarding low-mass dimuons, improving their measurement with the MFT is important since they can provide an important insight into the bulk properties and the evolution of the QCD matter formed in Pb--Pb collisions. The MFT will reduce the combinatorial
background from the semi-muonic decays from kaons and pions and will improve the mass resolution, thanks to the precise measurement of the opening angle of muon pairs \cite{ALICEMFTaddendum} \cite{lowmass}.

\section{Conclusions}

During the LS2, MCH and MID have been upgraded with new front-end and readout electronics to cope with the higher interaction rates in Run 3. The SAMPA electronics for the MCH is a result of an R$\&$D program developed in synergy with the TPC. Both the MCH and MID detectors are now under commissioning with first pp collision data. The installaiton of the MFT will significantly improve the muon physics program for Run 3 and Run 4 in the forward rapidity region. The ALPIDE chips used by the MFT (and ITS) are a result of an intense R$\&$D program. The MFT has been installed during the LS2, and is now under commissioning.  
All the components for the MS data readout chain have also successfully been tested and installed.

\end{document}